\documentclass[11pt,a4paper]{article}
\usepackage[utf8]{inputenc}
\usepackage[T1]{fontenc}
\usepackage{amsmath,amssymb}
\usepackage{graphicx}
\usepackage{booktabs}
\usepackage{hyperref}
\usepackage[margin=2.5cm]{geometry}

\title{Whisper-RIR-Mega: A Paired Clean-Reverberant Speech Benchmark for ASR Robustness to Room Acoustics}
\author{
Mandip Goswami\\
\texttt{Acoustics Researcher, WA, USA}\\
\texttt{mandipgoswami@gmail.com}\\
}
\date{February, 2026}

\begin{document}
\maketitle

\begin{abstract}
We introduce Whisper-RIR-Mega, a benchmark dataset of paired clean and reverberant speech for evaluating automatic speech recognition (ASR) robustness to room acoustics. Each sample pairs a clean LibriSpeech utterance with the same utterance convolved with a real room impulse response from the RIR-Mega corpus, with stratified splits by reverberation time (RT60) and direct-to-reverberant ratio (DRR). We evaluate five Whisper models (tiny through large-v3) on 1600 test samples and report word error rate (WER) and character error rate (CER) under clean and reverberant conditions. Reverberation consistently degrades performance across all model sizes; the reverb penalty in WER ranges from 2.31 to 15.50 percentage points depending on the model. Whisper-large-v3 shows the smallest penalty; Whisper-tiny shows the largest. We release the dataset, evaluation code, and baseline results to support reproducible research on robust ASR.
\end{abstract}

\section{Introduction}
\label{sec:intro}

Automatic speech recognition (ASR) systems are often trained and evaluated on relatively clean, close-talk recordings. In real environments, speech is captured in rooms where reflections and reverberation alter the signal and can degrade recognition accuracy. Evaluating ASR under controlled reverberant conditions is necessary to understand robustness and to drive progress in acoustic modeling and dereverberation.

Existing benchmarks for reverberant speech either lack paired clean references, use synthetic or limited RIR sets, or do not stratify by acoustic measures such as RT60 and DRR. We present Whisper-RIR-Mega, a benchmark that pairs each clean utterance with a reverberant version produced by convolving with a single real room impulse response (RIR) from the RIR-Mega corpus. Splits are stratified by RT60 (or DRR) when metadata is available, so the test set is balanced across acoustic conditions. This design allows direct comparison of clean versus reverberant WER and a clear reverb penalty (delta WER) per model.

We run baseline evaluations using five OpenAI Whisper models (tiny, base, small, medium, large-v3) on 1600 test samples. All models show higher WER and CER under reverberant speech; the smallest model (tiny) exhibits the largest reverb penalty in WER (15.50 percentage points), while large-v3 shows the smallest (2.31 percentage points). We release the dataset on Hugging Face, along with evaluation code and a leaderboard, to encourage reproducible benchmarks and further work on robust ASR.

\section{Related Work}
\label{sec:related}

Reverberation robust ASR has been addressed via multi-condition training, dereverberation front-ends, and end-to-end systems. Benchmark datasets include the REVERB challenge \cite{kinoshita2016reverb}, CHiME \cite{barker2015chime3}, and others that provide simulated or real reverberant speech. LibriSpeech \cite{panayotov2015librispeech} is widely used as a clean-speech source. RIR-Mega \cite{rirmega} is a large-scale simulated RIR dataset with machine-friendly metadata (RT60, DRR, C50); RIR-Mega-Speech \cite{rirmegaspeech} builds a reverberant speech corpus from LibriSpeech and RIR-Mega with per-file acoustic annotations and reproducible evaluation. Whisper \cite{radford2022whisper} provides strong open ASR baselines. Our benchmark combines LibriSpeech \cite{panayotov2015librispeech} and RIR-Mega \cite{rirmega} in a paired clean-reverb design with stratified splits and public baselines, complementary to \cite{rirmegaspeech}.

\section{Dataset Construction}
\label{sec:dataset}

\subsection{Speech and RIR Sources}
We use LibriSpeech test-clean \cite{panayotov2015librispeech} (16\,kHz) as the speech source. Each utterance is paired with a single room impulse response from the RIR-Mega dataset. RIR-Mega provides measured RIRs with metadata including RT60 (reverberation time) and DRR (direct-to-reverberant ratio). We use the train split of RIR-Mega and sample one RIR per utterance. When RT60 (or DRR) metadata is available, we stratify sampling across quantile bins so that the resulting dataset is balanced across acoustic conditions.

\subsection{Signal Processing}
For each utterance we convolve the clean waveform with the selected RIR at 16\,kHz. RIR energy is normalized before convolution, and the output is peak-normalized. No background noise is added. The clean and reverberant signals are stored as 16\,kHz FLAC. Each sample is identified by a unique \texttt{sample\_id} (derived from the LibriSpeech ID and the RIR assignment) and carries the reference transcript (\texttt{text\_ref}) and, when available, RIR metadata fields (e.g.\ \texttt{rir\_RT60\_T30\_s}, \texttt{rir\_DRR\_dB}).

\subsection{Splits}
We build 2000 paired samples in total. A deterministic split assigns 20\% to validation and 80\% to test (no training split in the default configuration). Assignment is stratified by RT60 quantiles when metadata exists, so that validation and test both reflect a similar distribution of room acoustics. The benchmark evaluation uses the test split (1600 samples).

\section{Experimental Setup}
\label{sec:setup}

\subsection{Models}
We evaluate five Whisper models: \texttt{openai/whisper-tiny}, \texttt{base}, \texttt{small}, \texttt{medium}, and \texttt{openai/whisper-large-v3}. Decoding uses beam size 5, best-of 5, temperature 0, and language set to English. All runs are performed on CPU with FP16 disabled for reproducibility.

\subsection{Metrics}
We report word error rate (WER) and character error rate (CER) using the \texttt{jiwer} library with standard normalization (lowercase, punctuation removal, whitespace collapse). For each model we compute mean WER and mean CER over the test set under two conditions: clean (original utterance) and reverb (convolved utterance). The reverb penalty is defined as the difference (reverb minus clean) in WER or CER.

\section{Results}
\label{sec:results}

Table~\ref{tab:leaderboard} reports mean WER and CER for each model on clean and reverberant test sets (1600 samples each condition). WER and CER use standard normalization (lowercase, punctuation removal). Clean WER ranges from 29.0\% (Whisper-large-v3) to 54.9\% (Whisper-tiny).

Reverberation increases WER for every model. The reverb penalty in WER (delta WER) is largest for Whisper-tiny (15.50 percentage points) and smallest for Whisper-large-v3 (2.31 percentage points). Whisper-medium and Whisper-small show intermediate penalties (5.94 and 7.44 percentage points). CER follows a similar pattern: reverb degrades performance, with the smallest penalty for Whisper-medium (0.48 pp) and the largest for Whisper-tiny (3.80 pp).

\begin{figure}[t]
\centering
\includegraphics[width=0.85\textwidth]{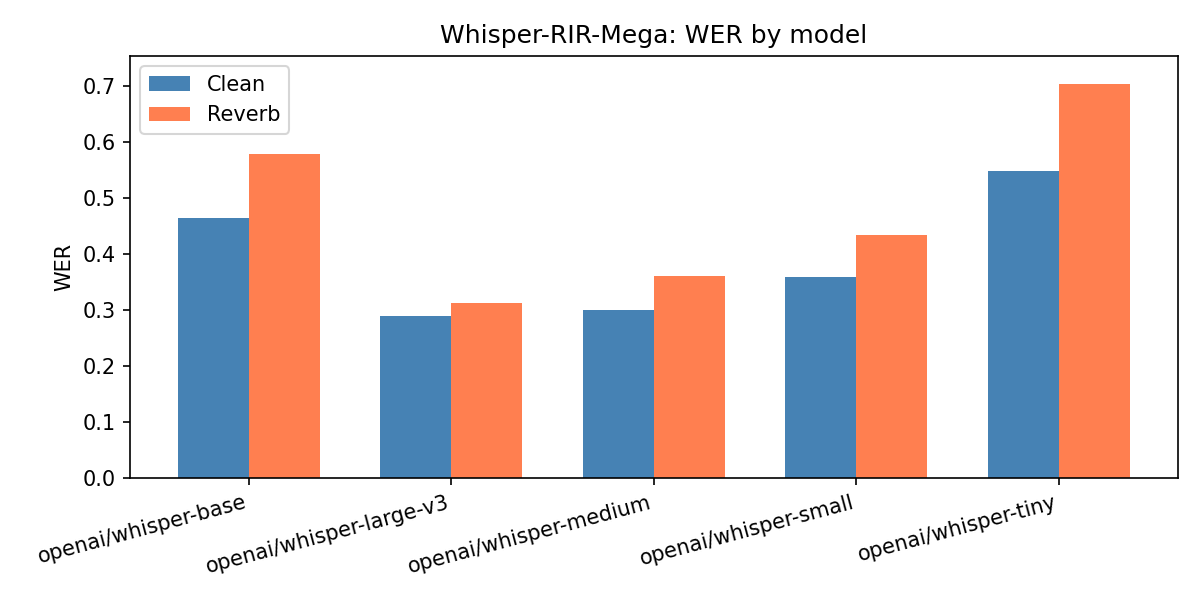}
\caption{WER by model: clean (blue) and reverberant (coral).}
\label{fig:leaderboard}
\end{figure}

Figure~\ref{fig:leaderboard} illustrates clean versus reverberant WER per model. When RIR metadata (RT60, DRR) is available in the dataset, the pipeline also produces WER versus RT60 bin and WER versus DRR bin (reverb condition); those plots are included in the repository and on the benchmark Space.

\begin{table}[t]
\centering
\caption{Mean WER (\%) and CER (\%) on the test set (1600 samples) for clean and reverberant conditions. $\Delta$WER and $\Delta$CER are reverb minus clean.}
\label{tab:leaderboard}
\begin{tabular}{lcccccc}
\toprule
Model & WER (clean) & WER (reverb) & $\Delta$WER & CER (clean) & CER (reverb) & $\Delta$CER \\
\midrule
Whisper-tiny   & 54.88 & 70.38 & 15.50 & 3.72 & 7.52 & 3.80 \\
Whisper-base  & 46.50 & 57.94 & 11.44 & 3.08 & 4.88 & 1.80 \\
Whisper-small & 35.88 & 43.31 & 7.44 & 2.05 & 3.04 & 0.99 \\
Whisper-medium & 30.06 & 36.00 & 5.94 & 1.90 & 2.38 & 0.48 \\
Whisper-large-v3 & 29.00 & 31.31 & 2.31 & 3.07 & 4.51 & 1.44 \\
\bottomrule
\end{tabular}
\end{table}

\section{Discussion}
\label{sec:discussion}

The results show that reverberation hurts ASR performance across all Whisper sizes, but the amount of degradation varies. Smaller models (tiny) are more sensitive to reverb in terms of WER delta; Whisper-large-v3 shows the smallest penalty. The relationship between model size and reverb sensitivity is monotonic: larger models exhibit smaller WER degradation. Robustness should be evaluated explicitly on benchmarks like Whisper-RIR-Mega.

The benchmark is limited to English (LibriSpeech) and to a single RIR per utterance. Extensions could include multiple RIRs per utterance, other languages, or additive noise. We encourage the community to use the dataset and report results on the public leaderboard.

\section{Conclusion}
\label{sec:conclusion}

We introduced Whisper-RIR-Mega, a paired clean-reverberant speech benchmark built from LibriSpeech and RIR-Mega, with stratified splits and baseline results for five Whisper models. Reverberation consistently increases WER and CER; the reverb penalty ranges from 2.31 to 15.50 percentage points in WER, with larger models showing greater robustness. The dataset, code, and leaderboard are publicly available to support reproducible research on robust ASR.

\section*{Data and Code Availability}
The dataset is hosted at \url{https://huggingface.co/datasets/mandipgoswami/whisper-rirmega-bench}. Evaluation code and reproduction instructions are in the repository \url{https://github.com/mandipgoswami/Whisper_RIRMega}. An interactive Space with the leaderboard and submission interface is at \url{https://huggingface.co/spaces/mandipgoswami/whisper-rirmega-benchmark}.

\end{document}